\documentclass[aps,physrev,superscriptaddress,amsmath,amssymb,
               nofootinbib,floatfix,reprint]{revtex4-2}
\usepackage{hyperref}
\usepackage{graphicx}
\usepackage{xcolor}
\usepackage[italicdiff]{physics}

\begin{document}

\title{Circuit growth and subspace read-out in quantum-selected configuration interaction: Separating the Hartree--Fock determinant}

\author{Ayaka Fujimoto}
\affiliation{Quantum Laboratory, Fujitsu Research, Fujitsu Limited., 4-1-1 Kamikodanaka, Nakahara, Kawasaki, Kanagawa 211-8588, Japan}
\email{fujimoto-ayaka@fujitsu.com}

\author{Norifumi Matsumoto}
\affiliation{Quantum Laboratory, Fujitsu Research, Fujitsu Limited., 4-1-1 Kamikodanaka, Nakahara, Kawasaki, Kanagawa 211-8588, Japan}

\author{Shota Kanasugi}
\affiliation{Quantum Laboratory, Fujitsu Research, Fujitsu Limited., 4-1-1 Kamikodanaka, Nakahara, Kawasaki, Kanagawa 211-8588, Japan}

\author{Kazunori Maruyama}
\affiliation{Quantum Laboratory, Fujitsu Research, Fujitsu Limited., 4-1-1 Kamikodanaka, Nakahara, Kawasaki, Kanagawa 211-8588, Japan}

\author{Hirotaka Oshima}
\affiliation{Quantum Laboratory, Fujitsu Research, Fujitsu Limited., 4-1-1 Kamikodanaka, Nakahara, Kawasaki, Kanagawa 211-8588, Japan}

\date{\today}

\begin{abstract}
Predicting many molecular properties requires accurate ground-state energies of molecular Hamiltonians. Quantum-selected configuration interaction (QSCI)
efficiently obtains such energies by measuring an input state to select important electronic
configurations and classically diagonalizing the Hamiltonian in the subspace
they span. ADAPT-QSCI, an adaptive extension of QSCI, improves the input state
through iterative circuit growth. For single-reference molecules, the sampling distribution is dominated by the Hartree--Fock (HF) configuration, so most measurements reproduce a configuration already known, and the remaining configurations become progressively more expensive to find. We reduce this redundancy by placing the HF configuration in the diagonalization subspace classically, before any measurement, rather than collecting it by sampling. This invalidates the operator-selection criterion of ADAPT-QSCI, and we derive a consistent replacement with energy and gradient in closed form, adding no query to the quantum device.
Because the proposed scheme and ADAPT-QSCI reach essentially the same subspace
at chemical accuracy, we separate the cost of constructing that subspace from the cost of reading it out. At the per-iteration shot budgets previously used for ADAPT-QSCI, the proposed scheme reaches chemical accuracy with approximately
$63\%$ and $43\%$ of the CNOT gates required by ADAPT-QSCI for H$_{4}$ and
H$_{6}$, respectively, while the total shot count falls to $44\%$ and $40\%$.
However, the relative advantage depends on the number of shots per iteration:
in the infinite-shot limit, the H$_{4}$ CNOT count increases to $125\%$ of
that of ADAPT-QSCI. These results demonstrate that the proposed scheme can improve the efficiency of QSCI calculations at appropriate per-iteration shot budgets, and identify the per-iteration shot budget as a key design variable governing resource savings in QSCI-based algorithms.
\end{abstract}

\maketitle

\section{Introduction}
\label{sec:intro}

Computing the ground-state energy of a molecular Hamiltonian is a central
task of electronic structure theory. Its exact solution in a finite basis,
full configuration interaction (FCI), requires a configuration space whose
dimension grows combinatorially with the numbers of electrons and orbitals.
Distributed implementations have reached $10^{12}$
determinants~\cite{Gao2024} and, more recently, $10^{15}$ in a distributed active-space framework~\cite{Shayit2025}.
These advances move the boundary of what is tractable without altering the scaling that sets it.
Quantum phase estimation~\cite{AspuruGuzik2005}  returns the ground-state energy to a prescribed accuracy, provided the initial state overlaps the ground state appreciably and the circuit runs on error-corrected hardware.
The variational quantum eigensolver was
proposed as an alternative for devices without quantum error correction, approximating the ground-state energy by minimizing the expectation value of the
Hamiltonian over a parameterized circuit~\cite{Peruzzo2014}. Estimating that expectation value is the main bottleneck. Here, the Hamiltonian is a sum of Pauli terms whose number grows steeply with the number of orbitals, each must be estimated to finite precision, and the shots required scale as the inverse square of the target accuracy. Resource analyses place the resulting count at of order $10^{10}$ circuit executions for chemical accuracy on molecules of practical interest~\cite{Gonthier2022}.

\begin{figure*}
\includegraphics[width=\textwidth]{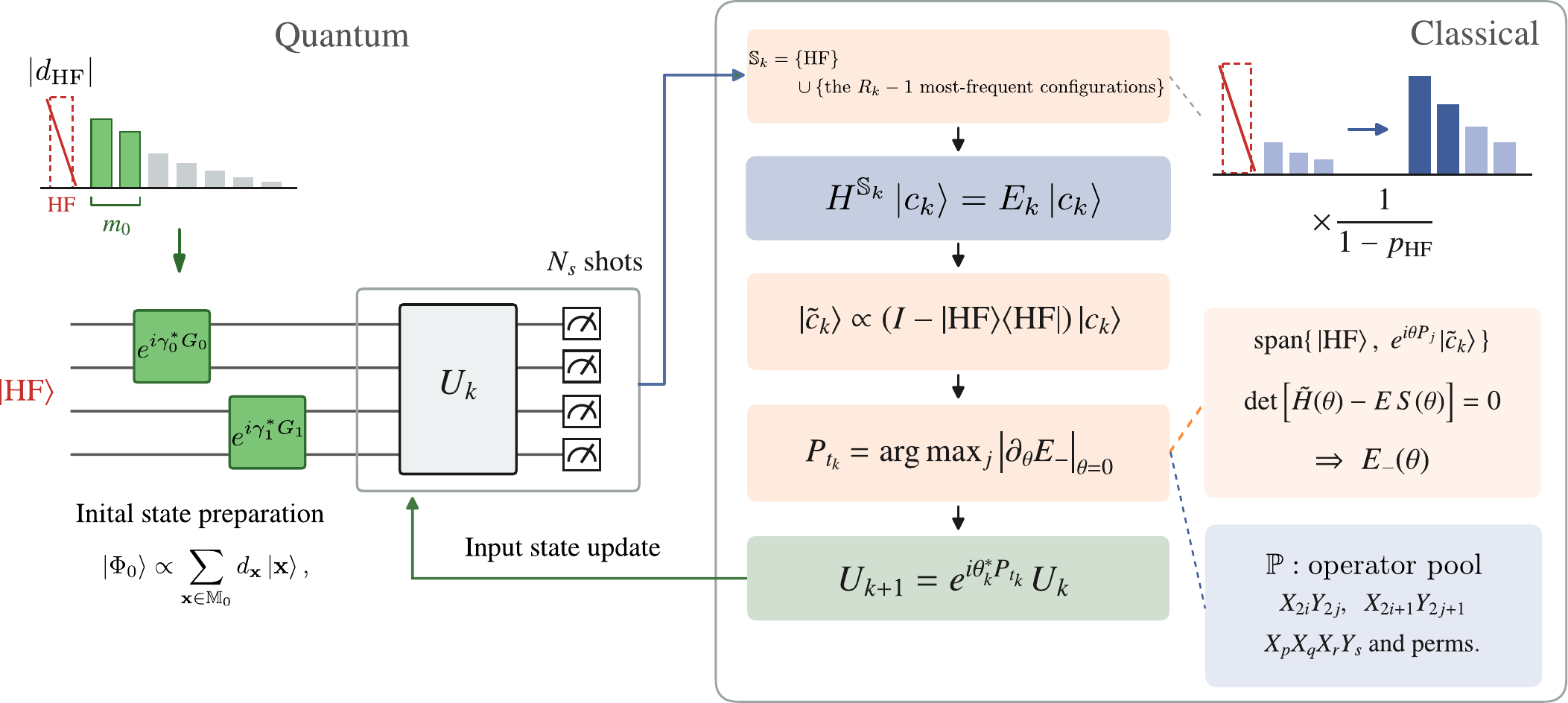}
\caption{Schematic of one iteration of the scheme. Orange boxes mark the
parts that differ from ADAPT-QSCI, blue boxes the parts it inherits
unchanged, and green boxes the two operations that append gates to the
quantum circuit, namely the preparation of the initial state and the update
of the circuit in the loop. Here $d_{\mathbf{x}}$ is the first-order
perturbation coefficient of configuration $\mathbf{x}$ and $\mathbb{M}_{0}$
the set of the $m_{0}$ configurations with the largest $\abs{d_{\mathbf{x}}}$
excluding the Hartree--Fock determinant, $G_{l}$ and $\gamma^{\ast}_{l}$ are
the generator and angle of the $l$th preparation rotation, and $U_{k}$ is the
circuit that prepares $\ket{\Phi_{k}}$.}
\label{fig:overview}
\end{figure*}

Quantum-selected configuration interaction (QSCI)~\cite{Kanno2026}, also known as sample-based quantum diagonalization (SQD)~\cite{RobledoMoreno2025}, avoids this by not
measuring an expectation value at all. An input state is
prepared and measured in the computational basis. Here, each outcome is an
electronic configuration, the configurations that appear span a subspace, and
the Hamiltonian is projected onto that subspace and diagonalized on a
classical computer~\cite{Kanno2026, RobledoMoreno2025}. No decomposition into Pauli terms is required, and the
accuracy is set by the subspace rather than by the statistics
of the measurement, which is used only to decide which configurations enter
it. QSCI shares with selected configuration interaction~\cite{Huron1973,Evangelista2014,Holmes2016,Tubman2016} the assumption that the ground state is carried by a small number of determinants,
and the two differ in how those determinants are found: selected
configuration interaction grows the set by a classical heuristic, ranking
candidate determinants by a perturbative estimate of the energy they would
bring, whereas QSCI takes whichever configurations the measurement returns. 

Constructing a good input state is the central design problem of QSCI. ADAPT-QSCI~\cite{Nakagawa2024} addresses
this by growing the state iteratively: the subspace
returned by QSCI supplies a classical eigenvector, the operators of a fixed pool are
ranked by the energy gradient evaluated on it, and the best one is appended
as a rotation. Other routes to the same problem take the input state from a
real-time evolution of the Hamiltonian~\cite{Mikkelsen2025,Sugisaki2025}. 

Across the QSCI family, the
efficiency with which sampling discovers new configurations has been
questioned~\cite{Reinholdt2025,vanRossum2026}. Part of that
inefficiency is structural. In a single-reference molecule, the sampling
distribution is dominated by one configuration: for the ADAPT-QSCI states considered here the Hartree--Fock determinant carries more than $0.94$ of the
weight at the iteration where chemical accuracy is reached, so that the great
majority of shots return a configuration that is already known. Concentration
of this kind makes the discovery of the remaining configurations
progressively more expensive~\cite{Reinholdt2025}, and the natural remedy,
spreading the weight more evenly, is obstructed by the fact that one does not
know in advance which configurations should receive it.

This work addresses the one configuration that is the exception. The Hartree--Fock determinant is fixed by the mean-field calculation that
defines the orbitals and costs nothing to write down, so there is no reason
to spend measurements on it. We therefore keep it out of the measured
distribution and supply it classically. The selection criterion has to be
rebuilt for this: the reference vector is stripped of its Hartree--Fock
component, which is restored in the objective as the lower root of a
two-by-two generalized eigenvalue problem, and the operators are ranked by
the gradient of that root. Figure~\ref{fig:overview} summarizes one iteration of the resulting scheme.

Removing the dominant outcome from the sampler renormalizes the probability of every remaining configuration, so the shots needed to collect them should fall by a factor set by the weight that was removed. On the two hydrogen chains studied here, H$_{4}$ and H$_{6}$, that factor is about twenty, and simulation recovers it to within a few per cent when the input state is held fixed. At the shot budgets of Ref.~\cite{Nakagawa2024} the scheme reaches chemical accuracy with $37\%$ fewer CNOT gates on H$_{4}$ and $57\%$ fewer on H$_{6}$, while reducing the total number of shots by factors of $2.25$ and $2.52$, respectively. Under the condition that the number of shots per iteration, a saving of this size in what a single iteration requires can be realized only as a reduction in the number of iterations, an integer that saturates once the sampler resolves the subspace. That is why a factor of twenty at fixed state becomes a factor of two or three for a complete run.
The sign of the comparison is set by the shot budget, and on each system there is a budget at which it changes.

The paper is organized as follows. Section~\ref{sec:setting} reviews QSCI and
ADAPT-QSCI and summarizes the notation, the two convergence criteria and the two
shot counts used throughout. Section~\ref{sec:res} collects the measures by
which the sampling distributions are described. Section~\ref{sec:scheme}
presents the scheme, gives the criterion and its gradient in closed form, and
predicts the reduction in the shot requirement that removing the
Hartree--Fock determinant should bring. Section~\ref{sec:num} reports the numerical validation, first for the
cost of building the subspace and then for the cost of reading it out.
Section~\ref{sec:open} concludes, separating what the experiments establish
from what they do not. Appendix~\ref{app:closedforms} gives the operator pool, the closed forms, a
verification of the analytic gradient, and the rule by which CNOT gates are
counted.

\section{Setting and notation}
\label{sec:setting}

In this section, we review QSCI and ADAPT-QSCI, and fix the notation used throughout (Table~\ref{tab:symbols}).
\subsection{Quantum-selected configuration interaction}
\label{sec:setting-qsci}

We consider an $n$-qubit Hamiltonian $H$ obtained from a fermionic
Hamiltonian through the Jordan--Wigner transformation~\cite{Jordan1928}, under
which each computational basis state $\ket{\mathbf{x}}$,
$\mathbf{x}\in\{0,1\}^{n}$, corresponds to one electronic configuration.
Configurations carrying the correct electron number $N_{e}$ and total $S_{z}$
are said to lie in the physical sector.

QSCI~\cite{Kanno2026} obtains an approximate ground-state energy by
diagonalizing $H$ in a subspace selected by measurement. A projective
measurement of an input state $\ket{\Phi}$ in the computational basis returns
$\mathbf{x}$ with probability
$P_{\Phi}(\mathbf{x})=\abs{\braket{\mathbf{x}}{\Phi}}^{2}$; one such
measurement is a shot, and $N_{s}$ shots give the observed frequency of each
configuration. Configurations observed with frequency below a threshold, taken to be $10^{-10}$ throughout this work, and
those outside the physical sector are discarded, and the $R$ most-frequent of
the remainder span a subspace $\mathbb{S}$ of dimension at most $R$. The
projected Hamiltonian
$H^{\mathbb{S}}$, with elements
$H^{\mathbb{S}}_{\mathbf{x}\mathbf{y}}=\mel{\mathbf{x}}{H}{\mathbf{y}}$, is
then diagonalized on a classical computer. The value of $R$ is fixed as in
Ref.~\cite{Nakagawa2024}, following the prescription of
Ref.~\cite{Kohda2022}; the values used here are given in
Sec.~\ref{sec:num}.

Neither classical step is expensive. Membership of the physical sector is
decided by $O(n)$ bit operations on a named $\mathbf{x}$, and
$H^{\mathbb{S}}_{\mathbf{x}\mathbf{y}}$ follows for any named pair in $O(n)$
from the Slater--Condon rules~\cite{Helgaker2000}.

\subsection{ADAPT-QSCI}
\label{sec:setting-adapt}

ADAPT-QSCI~\cite{Nakagawa2024} builds the input state for QSCI iteratively,
borrowing the operator-by-operator growth of ADAPT-VQE~\cite{Grimsley2019,Tang2021}, and proceeds as follows.

\begin{enumerate}
\item \emph{Initialization.}
  Define an operator pool $\mathbb{P}=\{P_{1},\cdots,P_{T}\}$ composed of
  single Pauli operators [Eq.~(S1) of Ref.~\cite{Nakagawa2024}, following
  qubit ADAPT-VQE~\cite{Tang2021}] and an initial state $\ket{\Phi_{0}}$,
  taken to be the Hartree--Fock state $\ket{\mathrm{HF}}$ in the runs reported below, as in Ref.~\cite{Nakagawa2024}. Set the number of iterations
  $k=0$.

\item \emph{Sampling and subspace.}
  Perform QSCI with the input state $\ket{\Phi_{k}}$, $N_{s}$ shots and
  maximum subspace dimension $R$. This generates the $R_{k}(\leq
  R)$-dimensional subspace $\mathbb{S}_{k}$, the projected Hamiltonian
  $H^{\mathbb{S}_{k}}$, its smallest eigenvalue $E_{k}$ and the associated
  eigenvector $\ket{c_{k}}$. If a series of $\{ E_k \}$ has converged, the algorithm stops.

\item \emph{Ranking.}
  Evaluate
  \begin{equation}
    h_{j}=\ev{i\comm{H}{P_{j}}}{c_{k}}
    \label{eq:hj}
  \end{equation}
  for all $P_{j}\in\mathbb{P}$ by classical computers. Equation~\eqref{eq:hj}
  is the derivative of
  $\ev{e^{-i\theta P_{j}}He^{i\theta P_{j}}}{c_{k}}$ at $\theta=0$.

\item \emph{Selection.}
  Choose the operator in the pool that has the largest $\abs{h_{j}}$ and
  denote it $P_{t_{k}}$.

\item \emph{Optimization.}
  Minimize
  \begin{equation}
    f_{k}(\theta)
    =\ev{e^{-i\theta P_{t_{k}}}He^{i\theta P_{t_{k}}}}{c_{k}}
    \label{eq:fk}
  \end{equation}
  with respect to $\theta$, and let $\theta^{\ast}_{k}$ be the minimizer.
  Because $P_{t_{k}}^{2}=I$, Eq.~\eqref{eq:fk} reduces to
  \begin{align}
  f_{k}(\theta)&=\cos^{2}\theta\,\ev{H}{c_{k}}\notag\\
  &\quad+\sin^{2}\theta\,\ev{P_{t_{k}}HP_{t_{k}}}{c_{k}}\notag\\
  &\quad+\sin\theta\cos\theta\,\ev{i\comm{H}{P_{t_{k}}}}{c_{k}} ,
  \label{eq:fkexpand}
  \end{align}
  a trigonometric function of $\theta$ whose three coefficients are obtained
  classically within $\mathbb{S}_{k}$, so that the minimizing angle follows in closed
  form.

\item \emph{Update.}
  Define the new input state for QSCI as
  \begin{equation}
    \ket{\Phi_{k+1}}=e^{i\theta^{\ast}_{k}P_{t_{k}}}\ket{\Phi_{k}} .
    \label{eq:adaptupdate}
  \end{equation}
  Set $k\leftarrow k+1$ and go back to step 2.
\end{enumerate}

Steps 3 to 5 are carried out entirely on a classical computer, using the
$R_{k}$-dimensional vector $\ket{c_{k}}$ rather than the
$2^{n}$-dimensional state $\ket{\Phi_{k}}$. The quantum device enters only
through the sampling of step 2.

\subsection{Stopping rule, evaluation criterion, and shot counts}
\label{sec:setting-conv}

Two distinct criteria appear in this work and are used for different
purposes. The first is that of Ref.~\cite{Nakagawa2024}, which stops the
iteration when
\begin{equation}
  \abs{E_{k}-E_{k-1}}<10^{-5}\ \mathrm{Ha} ,
  \label{eq:convorig}
\end{equation}
and which we use only to reproduce the shot counts reported there
(Sec.~\ref{sec:num}). The second is
\begin{equation}
  k^{\ast}=\min\qty{k:\ \abs{E_{k}-E_{\mathrm{FCI}}}\leq 1.6\ \mathrm{mHa}} ,
  \label{eq:kstar}
\end{equation}
the first iteration at which chemical accuracy, $1.6$mHa($\approx 1$kcal/mol), is reached. Since
Eq.~\eqref{eq:kstar} refers to $E_{\mathrm{FCI}}$, it is a measure used to
evaluate a completed run and not a rule by which a run can be stopped. It fixes the objective of the comparisons below, namely the cost of first reaching chemical accuracy, which is not the same as minimizing the total of gates and shots over the whole run.

We likewise distinguish two shot counts. The number of shots per iteration is
$N_{s}$, and is held fixed across methods and iterations; the total consumed
by a run is
\begin{equation}
  N_{\mathrm{tot}}=N_{s}\,(k^{\ast}+1) .
  \label{eq:ntot}
\end{equation}

\begin{table*}
\caption{Symbols used throughout. Quantities in the lower block are
evaluation measures: they refer to the exact ground state and are not
available to either method during a run.}
\label{tab:symbols}
\begin{ruledtabular}
\begin{tabular}{llc}
Symbol & Meaning & Defined in \\
\colrule
$n$ & number of qubits, equal to the number of spin orbitals
  & Sec.~\ref{sec:setting-qsci} \\
$\ket{\mathbf{x}}$ & computational basis state, one electronic configuration
  & Sec.~\ref{sec:setting-qsci} \\
$P_{\Phi}(\mathbf{x})$ & probability of obtaining $\mathbf{x}$ in one shot on
  $\ket{\Phi}$ & Sec.~\ref{sec:setting-qsci} \\
$N_{s}$, $N_{\mathrm{tot}}$ & shots per iteration; total shots,
  Eq.~\eqref{eq:ntot} & Sec.~\ref{sec:setting-conv} \\
$R$, $R_{k}$ & maximum subspace dimension; dimension obtained at iteration
  $k$ & Sec.~\ref{sec:setting-qsci} \\
$\mathbb{P}$, $P_{j}$, $P_{t_{k}}$ & operator pool; one of its elements; the
  element selected at iteration $k$ & Sec.~\ref{sec:setting-adapt} \\
$\theta^{\ast}_{k}$ & optimal rotation angle at iteration $k$
  & Sec.~\ref{sec:setting-adapt} \\
$\ket{\Phi_{k}}$ & input state prepared by the quantum circuit at iteration
  $k$ & Sec.~\ref{sec:setting-adapt} \\
$\mathbb{S}_{k}$, $H^{\mathbb{S}_{k}}$ & selected subspace; Hamiltonian
  projected onto it & Sec.~\ref{sec:setting-qsci} \\
$E_{k}$, $\ket{c_{k}}$ & smallest eigenvalue of $H^{\mathbb{S}_{k}}$ and its
  eigenvector & Sec.~\ref{sec:setting-adapt} \\
$h_{j}$ & derivative at $\theta=0$ of the energy of the reference rotated by
  $e^{i\theta P_{j}}$; ranks the pool in ADAPT-QSCI, Eq.~\eqref{eq:hj}
  & Sec.~\ref{sec:setting-adapt} \\
$\ket{\tilde{c}_{k}}$ & normalized non-Hartree--Fock part of $\ket{c_{k}}$,
  Eq.~\eqref{eq:ctilde} & Sec.~\ref{sec:scheme-steps} \\
$m_{0}$ & number of configurations in $\ket{\Phi_{0}}$, which excludes
  $\ket{\mathrm{HF}}$ & Sec.~\ref{sec:scheme-steps} \\
$p_{\mathrm{HF}}$ & probability that a shot returns $\ket{\mathrm{HF}}$
  & Sec.~\ref{sec:scheme-prediction} \\
$r_{\mathrm{pred}}$ & predicted reduction in the shot requirement on removing
  $\ket{\mathrm{HF}}$ from the sampling distribution, Eq.~\eqref{eq:rpred}
  & Sec.~\ref{sec:scheme-prediction} \\
$w_{\mathrm{phys}}$ & weight of the physical sector, Eq.~\eqref{eq:wphys}
  & Sec.~\ref{sec:res} \\
$\alpha_{R}$, $\beta_{R}$, $s_{\mathrm{eff}}$ & concentration of the sampling
  distribution, Eqs.~\eqref{eq:alphabeta} and \eqref{eq:seff}
  & Sec.~\ref{sec:res-sampler} \\
\colrule
$k^{\ast}$ & first iteration reaching chemical accuracy,
  Eq.~\eqref{eq:kstar} & Sec.~\ref{sec:setting-conv} \\
$\alpha^{\mathrm{GS}}$, $\mathbb{T}_{R}$ & weight of
  $\ket{\psi_{\mathrm{GS}}}$ recovered by a set; the $R$ configurations of
  largest weight in $\ket{\psi_{\mathrm{GS}}}$ & Sec.~\ref{sec:res-gs} \\
$p_{\mathrm{chem}}$ & probability that a single run reaches chemical accuracy
  & Sec.~\ref{sec:res-gs} \\
\end{tabular}
\end{ruledtabular}
\end{table*}

\section{Resource measures for the sampling distribution}
\label{sec:res}

We characterize the sampling distributions of Sec.~\ref{sec:num} using the
concentration measures introduced in Ref.~\cite{RobledoMoreno2025}.

Postselection onto the physical sector
(Sec.~\ref{sec:setting-qsci}) retains a fraction
\begin{equation}
  w_{\mathrm{phys}}=\sum_{\mathbf{x}}{}^{\prime}\,P_{\Phi}(\mathbf{x})
  \label{eq:wphys}
\end{equation}
of the shots, where the primed sum runs over configurations with the correct
electron number and total $S_{z}$. Below, $P_{\Phi}$ denotes the
post-selected distribution, renormalized to unity.

\subsection{Concentration of the sampling distribution}
\label{sec:res-sampler}

Let the configurations be labeled in order of decreasing probability,
$P_{1}\geq P_{2}\geq\cdots$. The concentration of $P_{\Phi}$ is characterized
by
\begin{equation}
  \alpha_{R}=\sum_{I=1}^{R}P_{I},
  \qquad
  \beta_{R}=P_{R},
  \label{eq:alphabeta}
\end{equation}
and by the participation number
\begin{equation}
  s_{\mathrm{eff}}
  =\frac{1}{\sum_{\mathbf{x}}P_{\Phi}(\mathbf{x})^{2}} .
  \label{eq:seff}
\end{equation}
Here $\alpha_{R}$ is the weight of the $R$ most probable configurations,
$\beta_{R}$ is the smallest of their probabilities, and $s_{\mathrm{eff}}$ is a
measure of how many configurations carry appreciable weight.

A configuration $\mathbf{x}$ is absent from $N_{s}$ shots with probability
$(1-P_{\Phi}(\mathbf{x}))^{N_{s}}$, so that the $R$ most probable
configurations are all observed except with probability at most
$R\,e^{-N_{s}\beta_{R}}$. Requiring this failure probability to be at most
$\eta$ therefore gives
\begin{equation}
  N_{s}\gtrsim\frac{\ln(R/\eta)}{\beta_{R}},
  \label{eq:shotbound}
\end{equation}
following the derivation of Ref.~\cite{RobledoMoreno2025}.
Equation~\eqref{eq:shotbound} is evaluated for the exact ground state in
Sec.~\ref{sec:num}.

\subsection{Overlap with the exact ground state}
\label{sec:res-gs}

The measures of Sec.~\ref{sec:res-sampler} refer to the sampling distribution
alone. Two further quantities refer to the exact ground state
$\ket{\psi_{\mathrm{GS}}}$ and are used for evaluation only. For a set of
configurations $\mathbb{S}$,
\begin{equation}
  \alpha^{\mathrm{GS}}(\mathbb{S})
  =\sum_{\mathbf{x}\in\mathbb{S}}
   \abs{\braket{\mathbf{x}}{\psi_{\mathrm{GS}}}}^{2}
  \label{eq:alphaGS}
\end{equation}
is the weight of $\ket{\psi_{\mathrm{GS}}}$ recovered by $\mathbb{S}$, and we define 
$\mathbb{T}_{R}$ as the set of the $R$ configurations of the physical
sector with the largest
$\abs{\braket{\mathbf{x}}{\psi_{\mathrm{GS}}}}^{2}$, a reference set
independent of the method and of $N_{s}$.

We further define $p_{\mathrm{chem}}(\ket{\Phi},N_{s})$ as the probability
that a single run --- drawing $N_{s}$ shots from $\ket{\Phi}$, forming the
subspace, and diagonalizing --- yields an energy within chemical accuracy of
$E_{\mathrm{FCI}}$. It is estimated from $400$ independent trials in
Sec.~\ref{sec:num-readout}.

\section{The Hartree--Fock separation scheme and its objective}
\label{sec:scheme}

This section introduces our proposed scheme. This scheme differs from ADAPT-QSCI in that the Hartree--Fock determinant is supplied classically rather than sampled. That change requires the criterion by which operators are selected to be rebuilt, and that is what most of this section derives.

\subsection{The scheme}
\label{sec:scheme-steps}

The scheme proceeds as follows, and Fig.~\ref{fig:overview} shows one
iteration.

\begin{enumerate}
\item \emph{Initialization.}
  Define an operator pool $\mathbb{P}=\{P_{1},\cdots,P_{T}\}$ composed of
  single Pauli operators and an initial state $\ket{\Phi_{0}}$. We prepared
  $\ket{\Phi_{0}}$ as a superposition of the $m_{0}$ configurations with the
  largest first-order perturbation coefficients, excluding
  $\ket{\mathrm{HF}}$ itself. The preparation
  circuit is described in Sec.~\ref{sec:scheme-init}. Set the number of
  iterations $k=0$.

\item \emph{Sampling and subspace.}
  Perform QSCI with the input state $\ket{\Phi_{k}}$, $N_{s}$ shots, and
  maximum subspace dimension $R_k$. After the postselection of
  Sec.~\ref{sec:setting-qsci}, the subspace is
  \begin{equation}
    \mathbb{S}_{k}=\{\mathrm{HF}\}\cup
    \qty{\text{the $R_k-1$ most-frequent configurations}},
    \label{eq:subspace}
  \end{equation}
  which is $R_{k}(\leq R)$-dimensional subspace. Diagonalizing the projected
  Hamiltonian $H^{\mathbb{S}_{k}}$ gives its smallest eigenvalue $E_{k}$ and
  the associated eigenvector $\ket{c_{k}}$. If a stopping condition fixed in
  advance is met (Sec.~\ref{sec:setting}), the algorithm stops.

\item \emph{Reference vector.}
  Take the non-Hartree--Fock part of $\ket{c_{k}}$, normalized,
  \begin{equation}
    \ket{\tilde{c}_{k}}
    =\frac{\qty(I-\dyad{\mathrm{HF}})\ket{c_{k}}}
          {\norm{\qty(I-\dyad{\mathrm{HF}})\ket{c_{k}}}},
    \label{eq:ctilde}
  \end{equation}
  so that $\braket{\mathrm{HF}}{\tilde{c}_{k}}=0$ by construction.

\item \emph{Ranking.}
  Evaluate the criterion on $\ket{\tilde{c}_{k}}$ for all
  $P_{j}\in\mathbb{P}$ by classical computers.

\item \emph{Selection.}
  Choose the operator in the pool that has the largest value and denote it
  $P_{t_{k}}$.

\item \emph{Update.}
  Determine the optimal angle $\theta^{\ast}_{k}$ for $P_{t_{k}}$ and define a
  new input state for QSCI as
  \begin{equation}
    \ket{\Phi_{k+1}}=e^{i\theta^{\ast}_{k}P_{t_{k}}}\ket{\Phi_{k}} .
    \label{eq:update}
  \end{equation}
  Set $k\leftarrow k+1$ and go back to step 2.
\end{enumerate}

The criterion by which the operators were ranked in step 4 is derived in
Secs.~\ref{sec:scheme-consistency}--\ref{sec:scheme-gradient}.

\subsection{Consistency of the selection criterion}
\label{sec:scheme-consistency}

The energy $E_{k}$ at iteration $k$ is the smallest eigenvalue of the
projected Hamiltonian $H^{\mathbb{S}_{k}}$, whose subspace contains
$\ket{\mathrm{HF}}$. The rotation $e^{i\theta P_{j}}$ considered in the
operator selection, on the other hand, acts on the state prepared by the
quantum circuit, $\ket{\Phi_{k}}$.

In ADAPT-QSCI every configuration in $\mathbb{S}_{k}$, including
$\ket{\mathrm{HF}}$, is obtained by sampling $\ket{\Phi_{k}}$. The eigenvector
$\ket{c_{k}}$ therefore lies entirely on the support of the state the circuit
prepared, and can be used as the reference vector as it stands.

In the present scheme, $\ket{\mathrm{HF}}$ enters $\mathbb{S}_{k}$ whether or
not it is observed, while the remaining configurations enter only by sampling
$\ket{\Phi_{k}}$ and number at most $R_k-1$ [Eq.~\eqref{eq:subspace}]. By
construction, $\ket{\tilde{c}_{k}}$, obtained from $\ket{c_{k}}$ by removing
the $\ket{\mathrm{HF}}$ component and renormalizing, is supported on those sampled configurations alone. We therefore took
$\ket{\tilde{c}_{k}}$ as the reference vector.

$E_{k}$ is, however, the smallest eigenvalue of a subspace that contains
$\ket{\mathrm{HF}}$. What was removed from the reference must be restored in
the objective. We therefore took as the objective the estimated ground-state
energy in the two-dimensional space spanned by $\ket{\mathrm{HF}}$ and the
rotated vector $e^{i\theta P_{j}}\ket{\tilde{c}_{k}}$. The elements of this
two-dimensional problem can be written in closed form.

\subsection{The two-dimensional generalized eigenvalue problem}
\label{sec:scheme-gevp}

Let $\tilde{H}(\theta)$ and $S(\theta)$ be the Hamiltonian and overlap
matrices of the two vectors $\ket{\mathrm{HF}}$ and
$e^{i\theta P_{j}}\ket{\tilde{c}_{k}}$, taken in that order. At $\theta=0$ one
has $S(0)=I$, since $\braket{\mathrm{HF}}{\tilde{c}_{k}}=0$. For
$\theta\neq0$ the rotation generates a Hartree--Fock component in
$e^{i\theta P_{j}}\ket{\tilde{c}_{k}}$ in general, so $S(\theta)\neq I$ and the objective
is the lower root of a generalized, not an ordinary, eigenvalue problem.

We take the pool to consist of Pauli strings with an odd number of $Y$
factors. The pool of Eq.~(S1) of Ref.~\cite{Nakagawa2024}, introduced for
qubit ADAPT-VQE~\cite{Tang2021} and used throughout this work, has exactly
one, so that $P_{j}$ acts on a computational basis state as
$P_{j}\ket{x}=i(-1)^{x_{y}}\ket{x\oplus m_{j}}$ for a mask $m_{j}$. 
Using $P_{j}^{2}=I$ and $e^{i\theta P_{j}}=I\cos\theta+i\sin\theta\,P_{j}$,
and with $H$ and $\ket{\tilde{c}_{k}}$ real in the computational basis, the
Hamiltonian matrix elements are
\begin{align}
  \tilde{H}_{11}&=\mel{\mathrm{HF}}{H}{\mathrm{HF}},\notag\\
  \tilde{H}_{12}(\theta)&=\mel{\mathrm{HF}}{He^{i\theta P_{j}}}{\tilde{c}_{k}}
    =v\cos\theta-r_{H,j}\sin\theta,\notag\\
  \tilde{H}_{22}(\theta)&=\ev{e^{-i\theta P_{j}}He^{i\theta P_{j}}}{\tilde{c}_{k}}
    \notag\\
  &=A\cos^{2}\theta+B_{j}\sin^{2}\theta+C_{j}\sin\theta\cos\theta,
  \label{eq:Hel}
\end{align}
and the overlap matrix elements are
\begin{align}
  S_{11}&=S_{22}=1,\notag\\
  S_{12}(\theta)&=\mel{\mathrm{HF}}{e^{i\theta P_{j}}}{\tilde{c}_{k}}
    =-r_{S,j}\sin\theta,
  \label{eq:Sel}
\end{align}
where
\begin{align}
  v&=\mel{\mathrm{HF}}{H}{\tilde{c}_{k}},\notag\\
  r_{H,j}&=-i\mel{\mathrm{HF}}{HP_{j}}{\tilde{c}_{k}},\notag\\
  r_{S,j}&=-i\mel{\mathrm{HF}}{P_{j}}{\tilde{c}_{k}}
  \label{eq:coupl}
\end{align}
are real (Appendix~\ref{app:closedforms}), and
\begin{align}
  A&=\ev{H}{\tilde{c}_{k}},\notag\\
  B_{j}&=\ev{P_{j}HP_{j}}{\tilde{c}_{k}},\notag\\
  C_{j}&=\ev{i\comm{H}{P_{j}}}{\tilde{c}_{k}}
  \label{eq:ABC}
\end{align}
carry the same definitions as in ADAPT-QSCI but are evaluated on
$\ket{\tilde{c}_{k}}$ rather than on $\ket{c_{k}}$. Every quantity in
Eqs.~\eqref{eq:Hel}--\eqref{eq:ABC} is a matrix element between named
determinants of $\mathbb{S}_{k}$ and is therefore obtained classically; the
criterion requires no additional queries to the quantum device.

Setting $\det\qty[\tilde{H}(\theta)-E\,S(\theta)]=0$ gives the quadratic
\begin{equation}
  a(\theta)E^{2}+b(\theta)E+c(\theta)=0
  \label{eq:quadratic}
\end{equation}
with
\begin{align}
  a(\theta)&=1-S_{12}(\theta)^{2},\notag\\
  b(\theta)&=-\qty[\tilde{H}_{11}+\tilde{H}_{22}(\theta)
    -2\tilde{H}_{12}(\theta)S_{12}(\theta)],
  \label{eq:abc}\\
  c(\theta)&=\tilde{H}_{11}\tilde{H}_{22}(\theta)
    -\tilde{H}_{12}(\theta)^{2}.\notag
\end{align}
The two-dimensional problem and its lower root are defined for any pool of
Hermitian operators with $P_{j}^{2}=I$; what the odd number of $Y$ factors
adds is that every quantity in Eqs.~\eqref{eq:Hel}--\eqref{eq:ABC} is real, so
that the coefficients in Eq.~\eqref{eq:abc} take the stated form.

\subsection{Analytic gradient}
\label{sec:scheme-gradient}

Substituting the lower root $E_{-}(\theta)$ into Eq.~\eqref{eq:quadratic}
gives an identity in $\theta$. Differentiating it and solving for
$\dv{E_{-}}{\theta}$ gives
\begin{equation}
  \dv{E_{-}}{\theta}
  =-\frac{a'(\theta)E_{-}^{2}+b'(\theta)E_{-}+c'(\theta)}
         {2a(\theta)E_{-}+b(\theta)} .
  \label{eq:implicit}
\end{equation}
At $\theta=0$, Eqs.~\eqref{eq:Hel} and \eqref{eq:Sel} give $S_{12}(0)=0$,
$\tilde{H}_{12}(0)=v$ and $\tilde{H}_{22}(0)=A$, hence
\begin{equation}
  a(0)=1,\quad b(0)=-(\tilde{H}_{11}+A),\quad c(0)=\tilde{H}_{11}A-v^{2},
  \label{eq:abc0}
\end{equation}
and, because $S_{12}(0)=0$ makes $a'(0)$ vanish,
\begin{equation}
  a'(0)=0,\quad
  b'(0)=-\qty(C_{j}+2vr_{S,j}),\quad
  c'(0)=\tilde{H}_{11}C_{j}+2vr_{H,j}.
  \label{eq:dabc0}
\end{equation}
We ranked the operators $P_{j}\in\mathbb{P}$ by
$\abs{\dv{E_{-}}{\theta}}_{\theta=0}$, obtained from
Eqs.~\eqref{eq:implicit}--\eqref{eq:dabc0} with $r_{H,j}$, $r_{S,j}$ and
$C_{j}$ evaluated for each $P_{j}$. The coefficient $B_{j}$ does not appear in
that expression, so scanning the whole pool costs
$O(\abs{\mathbb{P}}\cdot R_{k})$. We evaluated $B_{j}$ only for $P_{t_{k}}$
and then minimized $E_{-}(\theta)$ over $\theta$ to obtain
$\theta^{\ast}_{k}$.

We checked Eqs.~\eqref{eq:implicit}--\eqref{eq:dabc0} against a central
difference in $\theta$ on H$_{4}$, forming the $2\times2$ problem explicitly
at $\theta=\pm10^{-5}$. The relative deviations lay between
$7.9\times10^{-10}$ and $3.1\times10^{-8}$ (Appendix~\ref{app:closedforms},
Table~\ref{tab:grad}).

\subsection{Predicted measurement cost}
\label{sec:scheme-prediction}

Let $p_{\mathrm{HF}}$ be the probability that a shot on a given input state
returns $\ket{\mathrm{HF}}$, after postselection onto the physical sector.
Deleting that outcome from the sampling distribution and supplying it
classically renormalizes the probability of every remaining configuration by
$1/(1-p_{\mathrm{HF}})$. The number of shots required to observe a given set
of configurations at fixed confidence is inversely proportional to their
probabilities, so the shot requirement is reduced by the factor
\begin{equation}
  r_{\mathrm{pred}}=\frac{1}{1-p_{\mathrm{HF}}} .
  \label{eq:rpred}
\end{equation}
Equation~\eqref{eq:rpred} is tested at fixed state in
Sec.~\ref{sec:num-readout} (Fig.~\ref{fig:mech}).

\subsection{Preparation of the input state}
\label{sec:scheme-init}

We prepared $\ket{\Phi_{0}}$ from $\ket{\mathrm{HF}}$ using only rotations
generated by the same pool, selected greedily without lookahead, the target
being the fidelity to the intended superposition. For $m_{0}=2$ two gates sufficed and
the fidelity was $1.0000$, so $\ket{\Phi_{0}}$ is prepared exactly in the
setting used throughout this work. 

In the runs reported below, the initial state of ADAPT-QSCI is taken to be the Hartree--Fock state, a product state
requiring no two-qubit gates, whereas the present scheme starts from a
superposition of $m_{0}$ determinants. We therefore counted the preparation
circuit on the same footing as the iteration gates, under the counting rule of
Appendix~\ref{app:closedforms}; for $m_{0}=2$ the preparation costs $12$
CNOTs.

\section{Numerical validation}
\label{sec:num}

We compare the present scheme with ADAPT-QSCI on the systems and at the
settings of Ref.~\cite{Nakagawa2024}. The two methods are found to reach the
same subspace, so that the comparison reduces to the cost of reaching it, and
we measure that cost along two axes: the circuit cost in
Sec.~\ref{sec:num-build} and the measurement cost in
Sec.~\ref{sec:num-readout}.

The systems are the hydrogen chains H$_{4}$ and H$_{6}$, a standard benchmark
for correlated methods~\cite{Motta2020}, at a bond length of $1.0$~\AA,
described in the STO-3G basis~\cite{Hehre1969} and mapped to $8$ and $12$
qubits. Held fixed throughout are the subspace dimension, $R=14$ for
H$_{4}$ and $R=119$ for H$_{6}$ under the prescription of
Ref.~\cite{Nakagawa2024}, and the operator pool, which contains $164$ and
$1050$ elements. The present scheme is run with $m_{0}=2$, the setting for
which the initial state is prepared exactly (Sec.~\ref{sec:scheme-init}).

The number of shots per iteration is not a fixed setting but the control
parameter of the comparison. Both sections below vary $N_{s}$ and treat the
noiseless limit $N_{s}\to\infty$ alongside finite budgets. Where a single
budget is needed we take that of Ref.~\cite{Nakagawa2024}, $N_{s}=10^{5}$ for
H$_{4}$ and $5\times10^{6}$ for H$_{6}$, which lies within the same order of
magnitude as Eq.~\eqref{eq:shotbound} evaluated for the exact ground state,
$1.7\times10^{5}$ and $1.4\times10^{6}$.

Run at those budgets under the convergence criterion of
Eq.~\eqref{eq:convorig}, our
ADAPT-QSCI baseline consumes $N_{\mathrm{tot}}=1.240\times10^{6}$ for
H$_{4}$ and $2.060\times10^{8}$ for H$_{6}$, within $1.6\%$ and $8.4\%$ of
the totals reported in Ref.~\cite{Nakagawa2024}. All comparisons below are
made against this baseline.

Because $k^{\ast}$ is defined by the accuracy attained
[Eq.~\eqref{eq:kstar}], the two methods stop at the same energy, but they
need not stop with the same subspace. In the noiseless limit they do: on
H$_{4}$ the subspace at $k^{\ast}$ is the same set of determinants for both,
and on H$_{6}$ the two sets share $86$ of the $94$ and $95$ determinants they
contain, with values of $\alpha^{\mathrm{GS}}$ agreeing to within $1.2 \times 10^{-4}$. We therefore report in Secs.~\ref{sec:num-build} and \ref{sec:num-readout} the cost of reaching a given subspace, not a difference in what is reached.
The number of gates required to reach it does differ between the two methods,
and we do not determine the origin of this difference. Quantities quoted for
the noiseless limit come from a single deterministic trajectory per method;
for finite budgets the number of trials is stated in the caption of the
corresponding figure.

\subsection{The cost of building the subspace}
\label{sec:num-build}

\begin{figure*}[htbp]
\includegraphics[width=\textwidth]{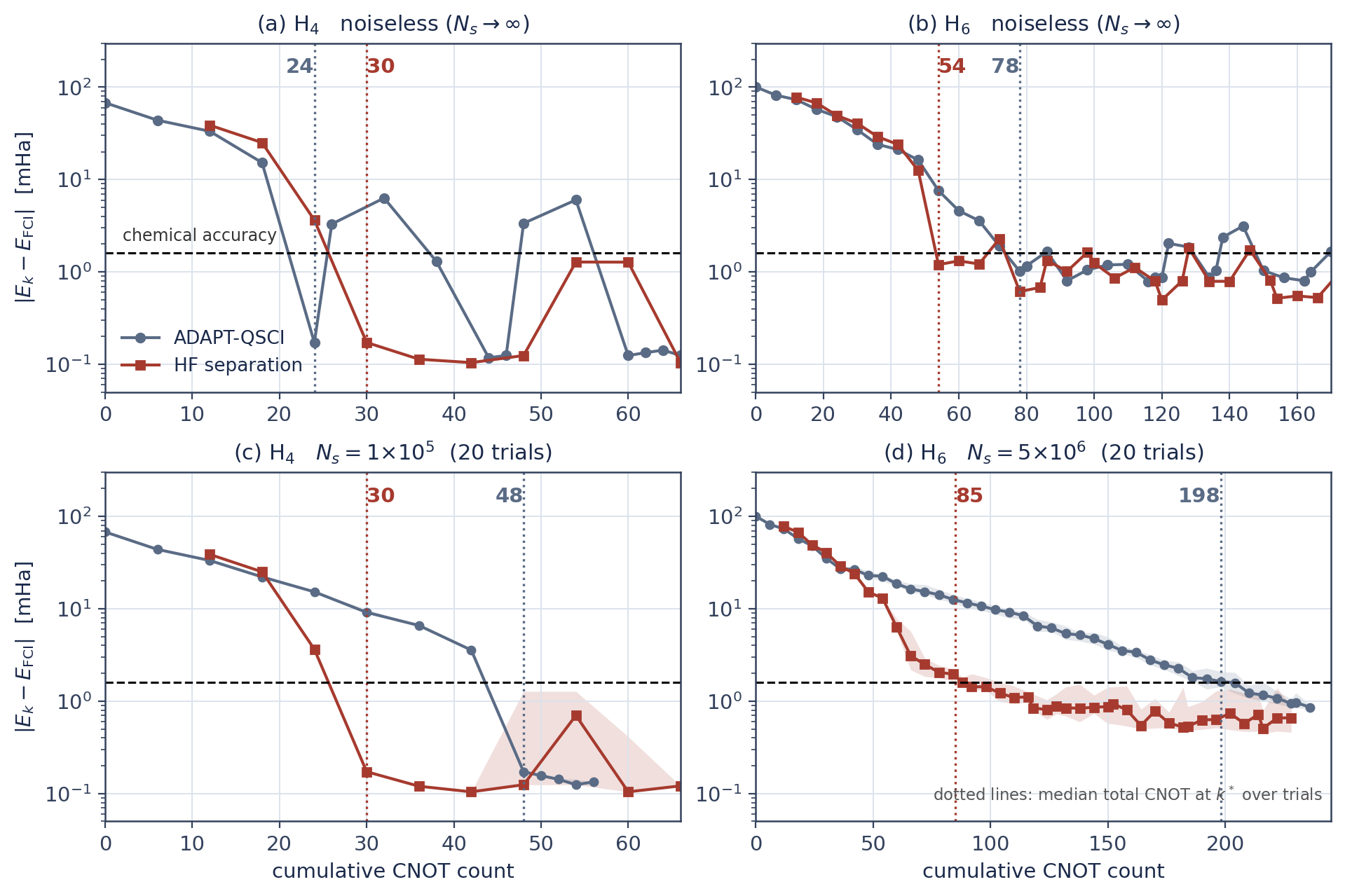}
\caption{Energy error against the cumulative CNOT count. Top row: noiseless
sampler, a single deterministic trajectory per method. Bottom row: the
budgets of Ref.~\cite{Nakagawa2024}, $N_{s}=10^{5}$ for H$_{4}$ and
$5\times10^{6}$ for H$_{6}$, median over $20$ trials with the interquartile
range shaded, each trial run to a fixed number of iterations, $12$ on
H$_{4}$ and $40$ on H$_{6}$, chosen to cover the point at which the baseline
reaches chemical accuracy; every trial reached it. Dashed horizontal lines:
chemical accuracy, $1.6$~mHa. Dotted vertical lines and labels: the total
CNOT count at $k^{\ast}$. The curves of the present scheme begin at $12$
CNOTs, the cost of preparing the $m_{0}=2$ initial state.}
\label{fig:traj}
\end{figure*}

The circuit cost of an input state is the number of two-qubit gates needed to
prepare it, two-qubit gates being the dominant cost and the dominant source
of error on noisy intermediate-scale quantum (NISQ) devices assumed here. We use the cumulative CNOT count, obtained
under the decomposition rule of Appendix~\ref{app:closedforms} and assuming
all-to-all connectivity, which is the resource measure of
Ref.~\cite{Nakagawa2024}. Figure~\ref{fig:traj} shows the energy error
$\abs{E_{k}-E_{\mathrm{FCI}}}$ against that count. The dotted vertical lines
mark the count at $k^{\ast}$, the first iteration at which the error falls
below chemical accuracy [Eq.~\eqref{eq:kstar}], and it is this count that we
compare. In the top row the sampler is noiseless, meaning that the exact
probabilities $P_{\Phi}$ are used in place of frequencies drawn from a finite
sample. The subspace at each iteration is then fixed by the state alone, and
each trajectory is a single deterministic run. In the bottom row the budget
is that of Ref.~\cite{Nakagawa2024}, $N_{s}=10^{5}$ for H$_{4}$ and
$5\times10^{6}$ for H$_{6}$, and the curves are medians over trials with the
interquartile ranges shaded.

The two methods do not start from the same state, and therefore not from the same place on the horizontal axis. ADAPT-QSCI begins at zero
CNOTs, its initial state being the Hartree--Fock product state, whereas the
present scheme begins at $12$, the cost of preparing $\ket{\Phi_{0}}$ from
$\ket{\mathrm{HF}}$ with the two rotations of Sec.~\ref{sec:scheme-init}. At
the budgets of Ref.~\cite{Nakagawa2024} that debt is recovered within the
first few iterations: chemical accuracy is reached with $30$ CNOTs on
H$_{4}$ against $48$ for ADAPT-QSCI, and with $85$ against $198$ on H$_{6}$,
reductions of $37\%$ and $57\%$.

The noiseless limit reverses the comparison on H$_{4}$, where the counts
become $24$ against $30$ and the present scheme costs $25\%$ more, and
reduces its margin on H$_{6}$, where they become $78$ against $54$. This is
to be expected: the limit supplies the sampler with an unlimited number of
shots, which is the resource the present scheme economizes, and it is
therefore the least favourable setting for it. The two rows of
Fig.~\ref{fig:traj} are thus the two ends of a range rather than two
independent results. Between them the comparison varies with the budget, and
on H$_{6}$ it reverses once more at the lowest budget of
Fig.~\ref{fig:perf}, where the present scheme needs more iterations than the
baseline. The sign of the comparison is set by the shot budget rather than by
the system.

The ratio of the two counts therefore factroized as $1.60 = 0.80 \times 2.00$ on H$_4$ and $2.33 = 1.44 \times 1.62$ on H$_6$, the noiseless value times what the budget of Ref.~\cite{Nakagawa2024} adds to it. The second factor is the subject of Sec.~\ref{sec:num-readout}; the noiseless value
is not accounted for there and is taken up in Sec.~\ref{sec:open}. On
H$_{6}$ the two are of comparable size.

\subsection{The cost of reading it out}
\label{sec:num-readout}

The sign of the comparison in Sec.~\ref{sec:num-build} is set by the shot
budget, and the budget enters through the number of shots needed to read a
subspace out of a given state. We measure that number under four
successively weaker conditions: with the state held fixed and only the
treatment of $\ket{\mathrm{HF}}$ changed; with the two states the methods
actually produce, compared at a matched iteration; with the states at which
they actually stop; and finally for the algorithm as a whole. Each condition
released transfers part of the advantage away from the read-out and into the
rest of the algorithm, and the four values are collected in
Table~\ref{tab:ladder}.

\begin{figure*}[htbp]
\includegraphics[width=0.68\textwidth]{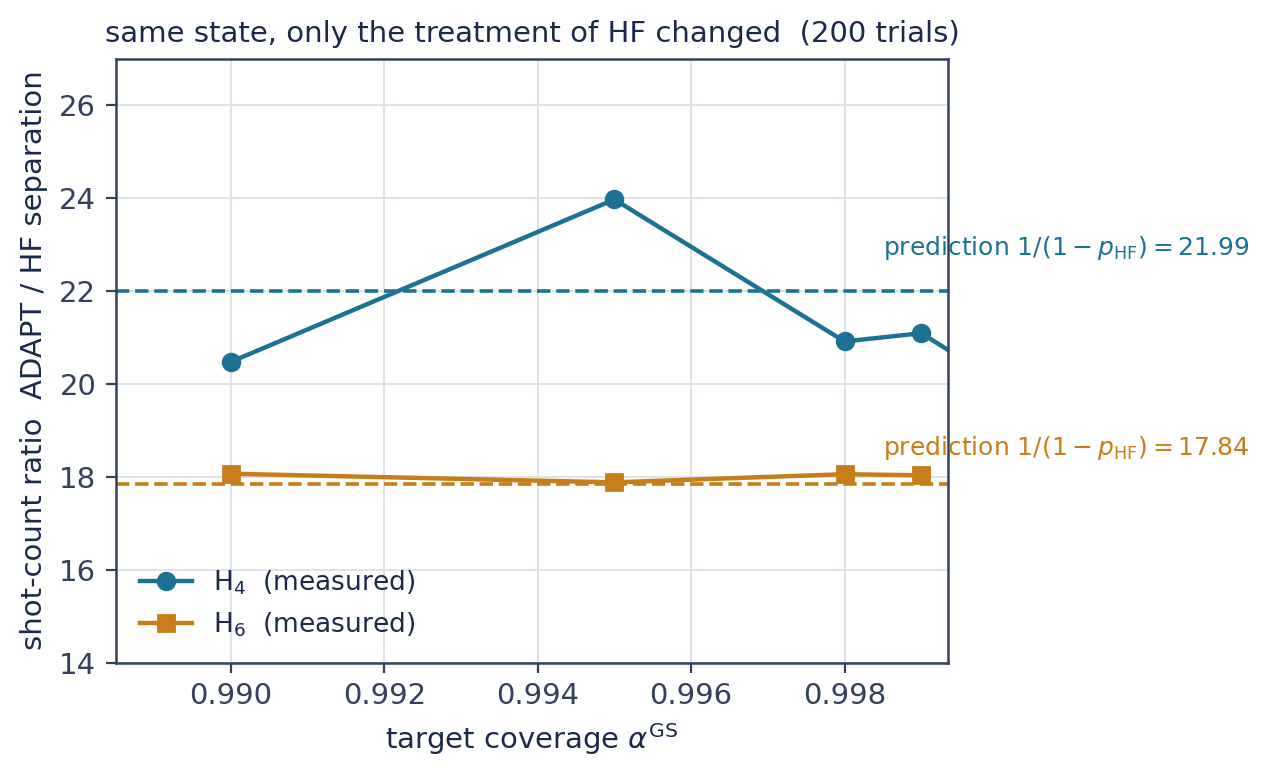}
\caption{Shot counts required to reach a target coverage
$\alpha^{\mathrm{GS}}$, for the state produced by ADAPT-QSCI at $k^{\ast}$
and for the same state with the Hartree--Fock outcome removed from the
sampler and supplied classically. The targets lie below
$\alpha^{\mathrm{GS}}(\mathbb{T}_{R})$, which is $0.999920$ for H$_{4}$ and
$0.999904$ for H$_{6}$. Dashed lines: the prediction $r_{\mathrm{pred}}$ of
Eq.~\eqref{eq:rpred}. $200$ trials.}
\label{fig:mech}
\end{figure*}

\subsubsection{The mechanism at fixed state}
\label{sec:num-mech}

The first comparison isolates the effect of Eq.~\eqref{eq:rpred} by removing
every other difference between the two methods. A single input state is
fixed, namely the state ADAPT-QSCI produces at its own $k^{\ast}$, which is
$k=4$ on H$_{4}$ and $k=13$ on H$_{6}$; these are the states for which
$p_{\mathrm{HF}}=0.954533$ and $0.943934$. Shots are drawn from that one state in two
ways. In the first, configurations are sampled from the post-selected
distribution $P_{\Phi}$ itself and the subspace is the $R$ most frequent of
them, as in ADAPT-QSCI. In the second, $\ket{\mathrm{HF}}$ is deleted from
$P_{\Phi}$ and the remaining probabilities are renormalized, the subspace
being $\ket{\mathrm{HF}}$ together with the $R-1$ most frequent of the rest,
as in the present scheme. The state, the value of $R$, the postselection and
the frequency threshold of Sec.~\ref{sec:setting-qsci} are common to the two,
so that the only difference left is whether $\ket{\mathrm{HF}}$ is obtained
from the sampler or supplied classically.

Each arm is characterized by the number of shots it needs to collect a
subspace of prescribed quality. We prescribe that quality as a target
coverage $\alpha^{\mathrm{GS}}$ [Eq.~\eqref{eq:alphaGS}] and, from a scan of
$N_{s}$ from $10^2$ to $10^{10.5}$, read off the shot count at which the coverage
averaged over $200$ trials first reaches it. The targets run from $0.990$ to
$0.9998$ on H$_{4}$ and to $0.999$ on H$_{6}$, in both cases below the
ceiling $\alpha^{\mathrm{GS}}(\mathbb{T}_{R})$ of $0.999920$ and $0.999904$
imposed by the reference set. Figure~\ref{fig:mech} plots the ratio of the
two shot counts against the target.

Equation~\eqref{eq:rpred} predicts both the value of that ratio and its
independence of the target, and the measurement bears out each. Its value,
$r_{\mathrm{pred}}=1/(1-p_{\mathrm{HF}})$, is $21.99$ on H$_{4}$ and $17.84$
on H$_{6}$ for the two states used here, drawn as dashed lines in
Fig.~\ref{fig:mech}; the measured ratios have geometric means $21.2$ and
$18.0$, so that the prediction holds to $-3.6\%$ and $+0.9\%$. It is
independent of the target because deleting $\ket{\mathrm{HF}}$ multiplies
every remaining probability by the same factor, whatever coverage is
demanded; accordingly the measured ratios stay within $20.4$--$24.0$ on
H$_{4}$ and $17.9$--$18.1$ on H$_{6}$, while the shot counts from which they
are formed grow by a factor of $670$ and $740$ across the range of targets.

\begin{figure*}[htbp]
\includegraphics[width=\textwidth]{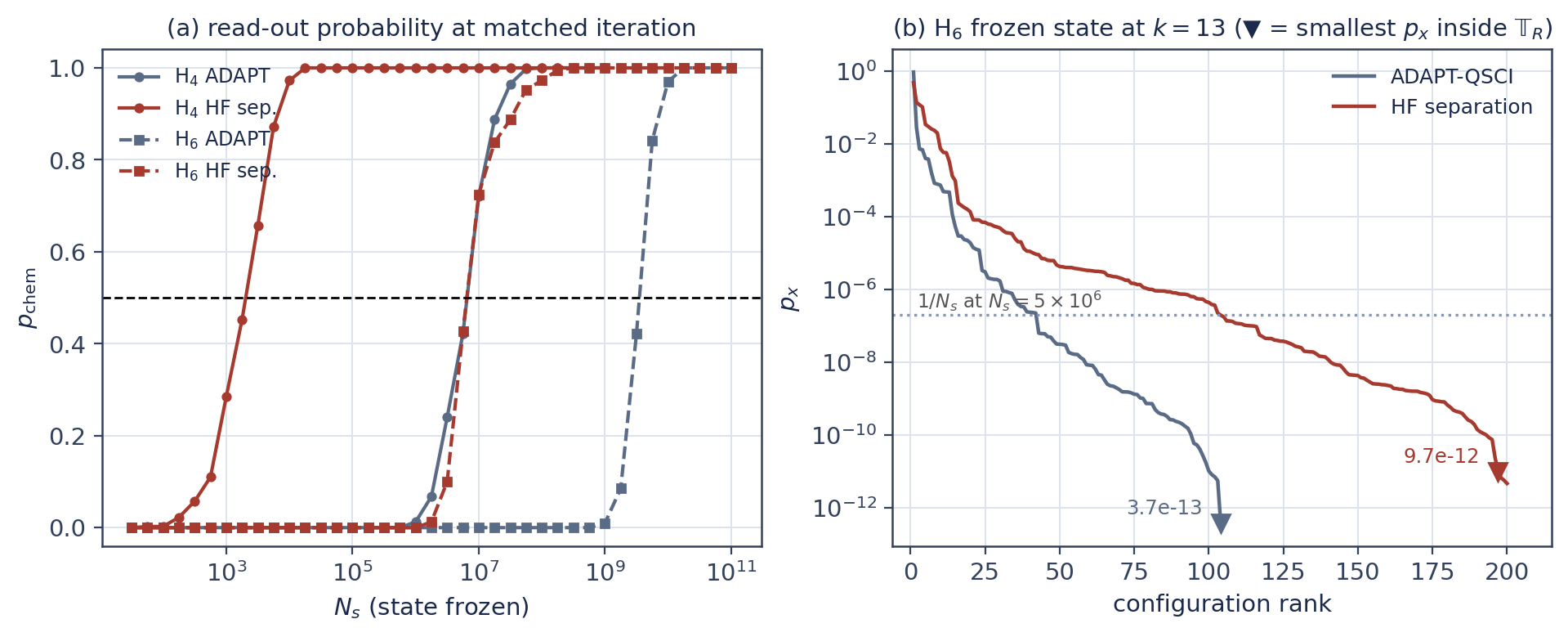}
\caption{(a) Probability $p_{\mathrm{chem}}$ that a single run achieves chemical accuracy, for the four frozen states compared at a matched
iteration, against the number of shots drawn from each. $400$ trials.
(b) Post-selected probabilities of the two H$_{6}$ states at $k=13$, ordered
by decreasing probability. Triangles: the smallest probability carried by
$\mathbb{T}_{R}$. Dotted line: $1/N_{s}$ at the budget of
Ref.~\cite{Nakagawa2024}.}
\label{fig:states}
\end{figure*}

\subsubsection{The states the two methods produce}
\label{sec:num-states}

The state is next allowed to differ, so that what is compared is no longer
two ways of sampling one state but the states the two methods actually build.
Figure~\ref{fig:states}(a) is based on four states held fixed: those of
ADAPT-QSCI and of the present scheme at $k=4$ on H$_{4}$ and at $k=13$ on
H$_{6}$. Taking both methods at the same iteration index serves a purpose
beyond symmetry: at these iterations the subspaces they build carry the same
weight of the exact ground state to four digits,
$\alpha^{\mathrm{GS}}(\mathbb{S}_{k})$ differing by $4.7\times10^{-5}$ on
H$_{4}$ and $1.2\times10^{-5}$ on H$_{6}$, so that the read-out costs below
are compared at equal subspace quality and not across states of different
usefulness.

Each of the four states is held fixed while $N_{s}$ is swept over the range
shown, and at every value of $N_{s}$ the following is repeated $400$ times:
$N_{s}$ shots are drawn from the post-selected distribution $P_{\Phi}$ of
that state; the subspace is formed from the outcomes by the rule belonging to
the method that produced the state, the $R$ most frequent configurations for
ADAPT-QSCI and $\ket{\mathrm{HF}}$ together with the $R-1$ most frequent of
the rest for the present scheme; and $H$ is diagonalized on it. The
frequency threshold of Sec.~\ref{sec:setting-qsci} is inert over this range,
a configuration seen even once being retained. The fraction of the $400$
subspaces whose lowest eigenvalue falls within chemical accuracy of
$E_{\mathrm{FCI}}$ is $p_{\mathrm{chem}}$ [Sec.~\ref{sec:res-gs}], and
it is this fraction that the four curves of Fig.~\ref{fig:states}(a) plot
against $N_{s}$. Each rises from zero to one over about two decades, and we
take the budget at which it passes $p_{\mathrm{chem}}=0.5$ as the read-out
cost of the state. Those budgets are $6.5\times10^{6}$ shots for the baseline
against $2.0\times10^{3}$ for the present scheme on H$_{4}$, and
$3.5\times10^{9}$ against $6.5\times10^{6}$ on H$_{6}$: ratios of $3216$ and
$544$, larger than the fixed-state ratios of Sec.~\ref{sec:num-mech} by
factors of $152$ and $30$.

Figure~\ref{fig:states}(b) traces that excess to the shape of the two
distributions. For the two H$_{6}$ states at $k=13$ it plots $P_{\Phi}$
itself, the exact post-selected probability of each configuration in the
frozen state, rather than frequencies drawn from any finite sample:
configurations outside the physical sector are removed and the remainder
renormalized, as they are before every subspace construction in this work,
and what is left is ordered by decreasing probability. The horizontal axis is
therefore a rank and not a label, and the two curves need not place the same
configuration at the same rank. The dotted line marks $1/N_{s}$ at the budget
of Ref.~\cite{Nakagawa2024}, below which a configuration is expected to
appear less than once in a run; at that budget $103$ configurations of the
present scheme lie above it against $42$ of the baseline, and of the $R=119$
members of the reference set $\mathbb{T}_{R}$ the corresponding counts are
$91$ and $41$. The triangles mark the smallest probability that
$\mathbb{T}_{R}$ carries in each state, $3.7\times10^{-13}$ for the baseline
against $9.7\times10^{-12}$ for the present scheme, this being the
configuration that sets the budget at which the whole of $\mathbb{T}_{R}$ is
collected. Summarizing the two profiles by the participation numbers of
Eq.~\eqref{eq:seff} gives $s_{\mathrm{eff}}=1.1$ and $3.6$. A flatter
profile is the cheaper of the two to read out, but we do not have a
quantitative model that turns these measures into the factors $152$ and $30$.

The matched iteration answers one question, namely which of two subspaces of
equal quality is cheaper to collect, while the operating points answer
another, since the two methods do not reach chemical accuracy at the same
iteration. Repeating the measurement with the state each method holds at its
own $k^{\ast}$ changes the H$_{4}$ ratio only from $3216$ to $3188$, the two
methods stopping one iteration apart there, but on H$_{6}$, where they stop
six iterations apart, it brings the ratio down from $544$ to $3.4$. The fall
is entirely on the side of the present scheme: the ADAPT-QSCI state is the
same in both comparisons, its $k^{\ast}$ being the $k=13$ used above, whereas
the state of the present scheme at $k^{\ast}=7$ costs $1.1\times10^{9}$ shots
to read out against $6.5\times10^{6}$ at $k=13$. The same two numbers show what continuing past $k^{\ast}$ would buy: six further iterations, $32$ more CNOTs, reduce the read-out budget by a factor of $165$. Where a run should stop if the total of gates and shots were to be minimized is a different question from the one $k^{\ast}$ answers, and we do not take it up. The advantage of
Fig.~\ref{fig:states}(a) is therefore not lost but spent: the present scheme
stops six iterations earlier, and a finite sampler must then work harder on
the less advanced state it stops with. Whether the exchange is favourable is
decided by the total shot count $N_{\mathrm{tot}}=N_{s}(k^{\ast}+1)$, which
carries both the shots per iteration and the number of iterations, and that
is the subject of Sec.~\ref{sec:num-total}.

\begin{figure*}[htbp]
\includegraphics[width=\textwidth]{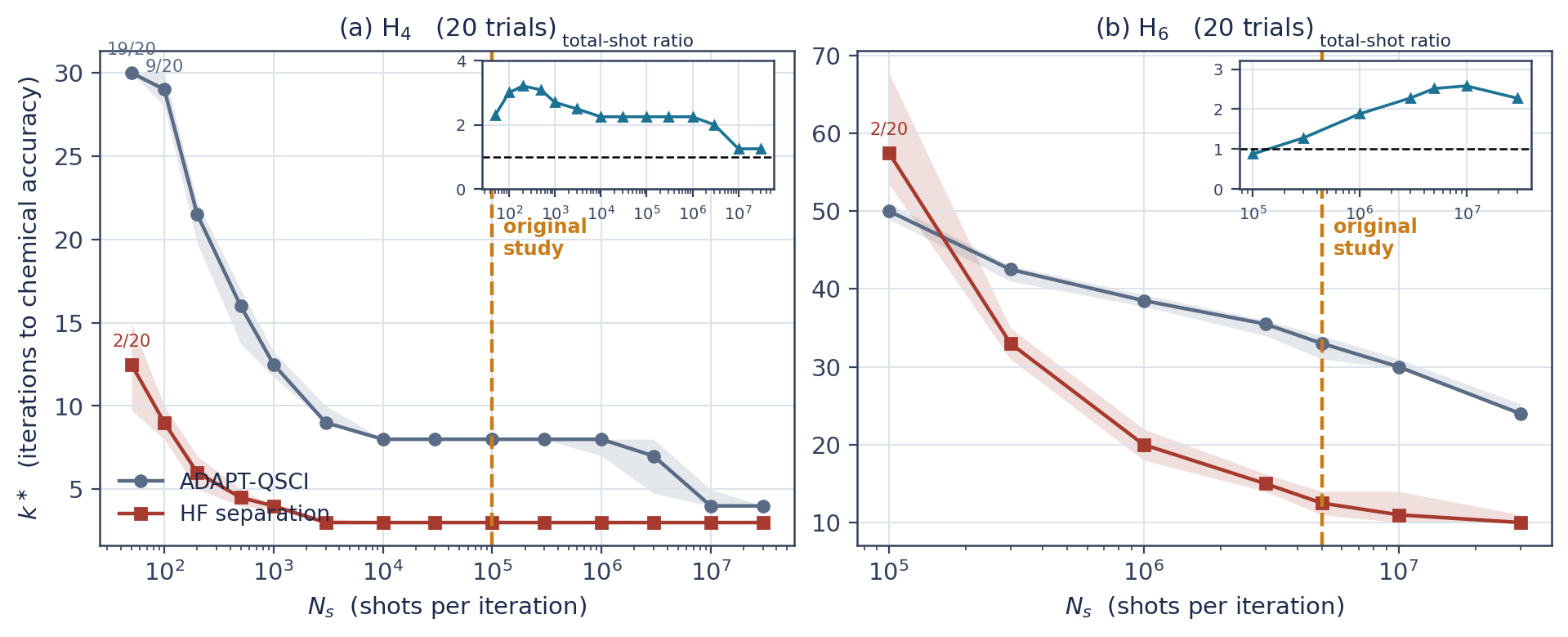}
\caption{Iterations $k^{\ast}$ to chemical accuracy against the number of
shots per iteration, median over $20$ trials with the interquartile range
shaded. Annotations give the number of trials that failed to converge, those
trials being entered at the iteration limit. Insets: the ratio of total shot
counts $N_{\mathrm{tot}}$, with unity marked. Dashed vertical lines: the
budgets of Ref.~\cite{Nakagawa2024}.}
\label{fig:perf}
\end{figure*}

\subsubsection{The algorithm as a whole}
\label{sec:num-total}

The condition released last is the freezing of the state. In the two
comparisons above a single input state was fixed and only the sampling of it
was varied, so that shot noise affected the read-out but not the state read
out. In a run of either algorithm it affects both: the subspace at each
iteration is built from a finite sample, the reference vector and hence the
operator and angle selected are inherited from it, and the state reached at
iteration $k$ therefore depends on every earlier sample. What follows
compares the two methods as complete algorithms under those conditions.

Figure~\ref{fig:perf} scans $N_{s}$ and, at each value, runs both methods to
convergence $20$ times from independent random seeds, recording $k^{\ast}$ in
each run. The median and interquartile range are plotted; runs that did not
reach chemical accuracy within the iteration limit are entered at that limit
and their number is annotated, which occurs only at the two smallest budgets
of H$_{4}$ and the smallest of H$_{6}$. The figure of merit is the total
number of shots a run consumes,
$N_{\mathrm{tot}}=N_{s}\left(k^{\ast}+1\right)$, the factor $k^{\ast}+1$
counting the sample taken at $k=0$ as well; the insets give the ratio of the
two totals, with unity marked. Since $N_{s}$ is held equal for the two
methods at each point of the scan, the ratio of the totals is also the ratio
of the iteration counts, and the comparison is one at fixed cost per
iteration rather than at fixed total cost.

The iteration counts behave differently in the two systems. On H$_{4}$ the
present scheme settles at $k^{\ast}=3$ from $N_{s}\approx3\times10^{3}$
upward, having then no further use for shots, while the baseline continues to
improve with the budget and reaches $k^{\ast}=4$ only near
$N_{s}=10^{7}$. On H$_{6}$ neither has saturated within the scan, the
baseline falling the more slowly of the two. The ratio of the totals runs
from $3.21$ down to $1.25$ on H$_{4}$ and from $0.87$ up to $2.58$ on
H$_{6}$, and at the budgets of Ref.~\cite{Nakagawa2024} it is $2.25$ and
$2.52$. The value below unity on H$_{6}$ is the smallest budget of the scan,
$N_{s}=10^{5}$, at which the present scheme is the more expensive of the two;
above $N_{s}\approx2\times10^{5}$ it is the cheaper.

\begin{table}
\caption{Ratios of shot counts, baseline over present scheme, for the four
comparisons of Sec.~\ref{sec:num-readout}, each holding fixed one condition
less than the one above it.}
\label{tab:ladder}
\begin{ruledtabular}
\begin{tabular}{lcc}
What is held fixed & H$_{4}$ & H$_{6}$ \\
\colrule
One state, Hartree--Fock treatment only & $21.2$ & $18.0$ \\
Matched iteration & $3216$ & $544$ \\
Each method at its own $k^{\ast}$ & $3188$ & $3.4$ \\
$N_{s}$ fixed, total shots & $1.25$--$3.21$ & $0.87$--$2.58$ \\
\end{tabular}
\end{ruledtabular}
\end{table}

Table~\ref{tab:ladder} collects the four comparisons, which do not form a
decreasing sequence. Reading a single state in two ways isolates the
mechanism of Eq.~\eqref{eq:rpred} and gives factors of $21.2$ and $18.0$;
comparing instead the states the two methods actually produce, at a matched
iteration, raises these to $3216$ and $544$, the largest entries in the
table, by the factors of $152$ and $30$ associated above with the shape of
the two distributions. Only from there does the advantage fall, and it falls
in two steps. The first is that the two methods stop at different iterations,
so that the state the present scheme reads out is not the one whose read-out
is cheap; this leaves H$_{4}$ unchanged and brings H$_{6}$ to $3.4$. The
second is that whatever survives can be collected only through $k^{\ast}$, an
integer that saturates once the sampler resolves the subspace, so that
further reductions in the cost per iteration buy nothing; for a complete run
the two totals then stand in ratios between $0.87$ and $3.21$. The same
$k^{\ast}$ sets the number of gates, which is why the circuit comparison of
Sec.~\ref{sec:num-build} inherits the dependence on $N_{s}$ seen there and
crosses over where it does.

\section{Conclusion}
\label{sec:open}

In this study we have modified quantum-selected configuration interaction so
that the Hartree--Fock determinant is kept out of the sampling distribution
and supplied classically. The reference vector on which operators are ranked is then no longer supported on the configurations the circuit prepares, so the selection criterion of ADAPT-QSCI does not apply as it stands. We derived one that does: the lower root of a two-by-two generalized eigenvalue problem spanned by the Hartree--Fock determinant and the rotated remainder of the subspace eigenvector. Both that root and its gradient are available in closed form and are evaluated classically, so the selection adds no query to the quantum device. Numerical
simulations on two hydrogen chains show that the scheme reaches chemical
accuracy with fewer CNOT gates and fewer measurement shots than ADAPT-QSCI at
the shot budgets of Ref.~\cite{Nakagawa2024}, that the shot requirement at
fixed state falls by the predicted factor $1/(1-p_{\mathrm{HF}})$, and that
the subspace obtained at chemical accuracy is essentially unchanged. The advantage is set by the shot budget rather than by the system, and on both molecules there is a budget at which the comparison changes sign.

Three points remain open. The number of gates needed to reach the same subspace in the
noiseless limit differs between the two methods, and we do not determine why.
The ratio between the states the methods actually produce exceeds the
prediction obtained at fixed state by factors of $152$ and $30$, for
which we have no quantitative model. The budget at which the comparison
changes sign, near $N_{s}=2\times10^{5}$ on H$_{6}$, where the ratio of total
shot counts in Fig.~\ref{fig:perf} crosses unity, is located but not
explained, and two systems are too few to say what fixes it. These three questions define the natural continuation of this work. 

The present study is also limited in scope. Both molecules are
single-reference and described in a minimal basis, and every comparison is
made at a fixed number of shots per iteration. Although $k^{\ast}$ measures the cost of first reaching chemical accuracy, the resource-optimal stopping point may occur at a later iteration because further state construction can reduce the subsequent read-out cost. Determining this optimum requires a joint cost model for state preparation and read-out and is left for future work. Finally, the sampling is free of device noise, and this is the most
consequential of the omissions. The two schemes reach chemical accuracy with
different gate counts, and the present scheme spends part of its budget on a
preparation circuit that ADAPT-QSCI does not need. Noise would therefore not
act on them equally, and a noisy simulation is needed to say in which
direction the comparison moves.

\section*{Acknowledgement}
The authors are profoundly indebted to Shintaro~Sato for his invaluable support in fostering a conducive research environment.
A.F. is grateful to Koki~Chinzei and Quoc~Hoan~Tran for helpful comments and discussions. 
A.F. acknowledges Masatoshi~Ishii for technical support in performing numerical simulations.

\appendix

\section{Closed forms and gradient verification}
\label{app:closedforms}

\subsection{The operator pool}
\label{app:pool}

The pool is that of Eq.~(S1) of Ref.~\cite{Nakagawa2024}, introduced for
qubit ADAPT-VQE~\cite{Tang2021}. Spin orbitals are ordered
$(0\uparrow),(0\downarrow),\ldots,(N_{\mathrm{MO}}-1\uparrow),
(N_{\mathrm{MO}}-1\downarrow)$ and labelled $0,\ldots,n-1$, so that even and
odd labels carry opposite spin. The pool then consists of the two-body
strings
\begin{equation}
  X_{2i}Y_{2j},\qquad X_{2i+1}Y_{2j+1},
  \qquad 0\leq i<j\leq n/2-1,
  \label{eq:pool2}
\end{equation}
and the four-body strings
\begin{equation}
  X_{p}X_{q}X_{r}Y_{s},\quad
  X_{p}X_{q}Y_{r}X_{s},\quad
  X_{p}Y_{q}X_{r}X_{s},\quad
  Y_{p}X_{q}X_{r}X_{s}
  \label{eq:pool4}
\end{equation}
with $0\leq p<q<r<s\leq n-1$ and $p+q+r+s$ even. Equation~\eqref{eq:pool2}
stands for generalized single excitations with the Jordan--Wigner strings
dropped, Eq.~\eqref{eq:pool4} for generalized double excitations, and the
parity condition mimics conservation of $S_{z}$.
Strings of $YX$ type are
omitted because the swap $X\leftrightarrow Y$ relates them to those of
Eq.~\eqref{eq:pool2}. Counting gives $12+152=164$ elements for $n=8$ and
$30+1020=1050$ for $n=12$, the values quoted in Sec.~\ref{sec:num}. Every
element contains exactly one $Y$ factor and has weight two or four, which is
what Secs.~\ref{app:parity} and \ref{app:cnot} rely on.

\subsection{Parity of the \texorpdfstring{$Y$}{Y} factors}
\label{app:parity}

In the computational basis $H$ is a real symmetric matrix, $\ket{c_{k}}$ and
with it $\ket{\tilde{c}_{k}}$ may be taken real, an eigenvector being defined
only up to a phase, and $\ket{\mathrm{HF}}$ is a basis vector. Since $X$ and
$Z$ are real matrices while $Y$ is $i$ times one, a Pauli string containing
$n_{Y}$ factors of $Y$ satisfies $P_{j}=i^{n_{Y}}M_{j}$ with $M_{j}$ real.
The two parities of $n_{Y}$ then have different consequences.

For $n_{Y}$ odd, $P_{j}$ is $i$ times a real matrix, and every quantity of
Eqs.~\eqref{eq:Hel}--\eqref{eq:ABC} is real. This holds for $v$ and $A$
because $H$ and the vectors are real, for $r_{H,j}$, $r_{S,j}$ and $C_{j}$
because the matrix elements they are built from are purely imaginary and the
prefactor removes the phase, and for $B_{j}$ because $P_{j}HP_{j}$ carries
two factors of $P_{j}$. This is what allows the coefficients of
Eq.~\eqref{eq:abc} to be written without absolute values or complex
conjugates, and Eq.~\eqref{eq:dabc0} with them.

For $n_{Y}$ even, $P_{j}$ is real and, being Hermitian, symmetric. Then
$\ev{HP_{j}}{\tilde{c}_{k}}$ and $\ev{P_{j}H}{\tilde{c}_{k}}$ are real and
equal to each other by transposition, so that $C_{j}$ vanishes. Writing
$\sigma=\mel{\mathrm{HF}}{P_{j}}{\tilde{c}_{k}}$ and
$\eta=\mel{\mathrm{HF}}{HP_{j}}{\tilde{c}_{k}}$, both real, one has
$S_{12}(\theta)=i\sigma\sin\theta$ and
$\tilde{H}_{12}(\theta)=v\cos\theta+i\eta\sin\theta$, so that
$\abs{S_{12}}^{2}$, $\abs{\tilde{H}_{12}}^{2}$ and
$\Re[\tilde{H}_{12}S_{12}^{*}]=\eta\sigma\sin^{2}\theta$ are even in
$\theta$. With $C_{j}=0$, so is
$\tilde{H}_{22}(\theta)=A\cos^{2}\theta+B_{j}\sin^{2}\theta$. Every
coefficient of Eq.~\eqref{eq:quadratic} is therefore even, hence so is
$E_{-}(\theta)$, and $\dv*{E_{-}}{\theta}$ vanishes at $\theta=0$. Such an
operator is never selected, which is why the pool may be taken to consist of
strings with an odd number of $Y$ factors at no cost.

\subsection{The lower root}
\label{app:lowerroot}

Since $\abs{S_{12}(\theta)}<1$ whenever the two vectors are linearly
independent, $a(\theta)>0$ and the lower root of Eq.~\eqref{eq:quadratic} is
\begin{equation}
  E_{-}(\theta)
  =\frac{-b(\theta)-\sqrt{b(\theta)^{2}-4a(\theta)c(\theta)}}{2a(\theta)} .
  \label{eq:Eminus}
\end{equation}
At $\theta=0$, Eq.~\eqref{eq:abc0} reduces this to the smaller eigenvalue of
the real symmetric matrix with diagonal $\tilde{H}_{11}$, $A$ and
off-diagonal $v$,
\begin{equation}
  E_{-}(0)=\frac{\tilde{H}_{11}+A}{2}
    -\sqrt{\qty(\frac{\tilde{H}_{11}-A}{2})^{2}+v^{2}} .
  \label{eq:Eminus0}
\end{equation}

\subsection{Verification of the analytic gradient}
\label{app:grad}
 
Table~\ref{tab:grad} compares Eqs.~\eqref{eq:implicit}--\eqref{eq:dabc0} with
a central difference of Eq.~\eqref{eq:Eminus} at $\theta=\pm10^{-5}$, the
$2\times2$ problem being formed explicitly in the full $2^{n}$-dimensional
space at each $\theta$. The first block is evaluated on the initial subspace
of H$_{4}$ with $m_{0}=2$. There $\ket{\tilde{c}_{0}}$ is supported on two
configurations only, and $r_{S,j}$, which is nonzero only when
$\mathbf{x}_{\mathrm{HF}}\oplus m_{j}$ lies in that support, vanishes for
every operator tested; the term $2vr_{S,j}$ of Eq.~\eqref{eq:dabc0} is then
not exercised. The second block therefore repeats the test at $k=2$, where
the subspace is nine-dimensional and $32$ operators of the pool have
$r_{S,j}\neq0$, using the four of largest $\abs{r_{S,j}}$.
 
Two entries of the first block are zero to machine precision. Both are
strings for which $C_{j}$, $r_{H,j}$ and $r_{S,j}$ all vanish on
$\ket{\tilde{c}_{0}}$, so that $b'(0)$ and $c'(0)$ vanish with them; they are
not instances of the even-$Y$ argument of Sec.~\ref{app:parity}, each
containing one $Y$ factor.
 
\begin{table}
\caption{Analytic gradient $\dv*{E_{-}}{\theta}$ at $\theta=0$ against a
central difference with step $10^{-5}$, on H$_{4}$ with $m_{0}=2$. Upper
block: the initial subspace, on which $r_{S,j}=0$ for every operator listed.
Lower block: the subspace at $k=2$, using the four operators of largest
$\abs{r_{S,j}}$. The relative deviation is $\abs{\text{analytic}-\text{central
difference}}/\abs{\text{analytic}}$; the two vanishing entries are reported
as absolute deviations.}
\label{tab:grad}
\begin{ruledtabular}
\begin{tabular}{lrrr}
Pauli string & $r_{S,j}$ & analytic & rel.\ dev. \\
\colrule
XIYIIIII & $0$ & $0$ & $0$ \\
IXIIIYII & $0$ & $-0.00137419$ & $3.1\times10^{-8}$ \\
YXIIIXXI & $0$ & $0$ & $0$ \\
IYXIXIIX & $0$ & $0.01482975$ & $1.8\times10^{-9}$ \\
IIIIYXXX & $0$ & $-0.03543426$ & $7.9\times10^{-10}$ \\
\colrule
IIYXXXII & $0.740876$ & $0.00559817$ & $1.5\times10^{-8}$ \\
IIXYXXII & $0.740876$ & $0.00559817$ & $1.5\times10^{-8}$ \\
IIXXYXII & $-0.740876$ & $-0.00559817$ & $1.5\times10^{-8}$ \\
IIXXXYII & $-0.740876$ & $-0.00559817$ & $1.5\times10^{-8}$ \\
\end{tabular}
\end{ruledtabular}
\end{table}

\subsection{Counting of CNOT gates}
\label{app:cnot}

A rotation $e^{i\theta P_{j}}$ generated by a Pauli string of weight $w$, the
number of its non-identity factors, is compiled as single-qubit basis changes
on the $w$ qubits, a ladder of $w-1$ CNOT gates, one single-qubit $Z$
rotation, and the reversed ladder. It therefore costs
\begin{equation}
  2(w-1)
  \label{eq:cnotrule}
\end{equation}
CNOT gates. All-to-all connectivity is assumed and single-qubit gates are not
counted, two-qubit gates being the dominant cost and the dominant source of
error on NISQ devices assumed here. Every operator selected up to $k^{\ast}$ in this work, in the
preparation circuit of Sec.~\ref{sec:scheme-init} as well as in the
iterations, is a weight-four string, so that each rotation costs six CNOT
gates and the totals quoted in Sec.~\ref{sec:num} are six times the number of
rotations applied.

\bibliography{qsci_v1}
\end{document}